\documentclass[useAMS,usenatbib]{mn2e}
\usepackage{graphicx}
\usepackage{lscape}
\newcommand{\lesssim}{\mathrel{\hbox{\rlap{\hbox{\lower4pt\hbox{$\sim$}}}\hbox{$<$}}}}
\newcommand{\gtrsim}{\mathrel{\hbox{\rlap{\hbox{\lower4pt\hbox{$\sim$}}}\hbox{$>$}}}}
\newcommand{\msun}{$M_{\odot}$}
\title[The diffusion-induced nova scenario]{The diffusion-induced nova
scenario.\\ CK Vul and PB 8 as possible observational counterparts}

\author[M. M. Miller Bertolami et al.]{M. M. Miller
Bertolami$^{1,2}$\thanks{E-mail: mmiller@fcaglp.unlp.edu.ar},
L. G. Althaus$^{1,2}$, C. Olano$^{1}$, and N. Jim\'enez$^{1,2}$\\
$^{1}$Facultad de Ciencias Astron\'omicas y Geof\'isicas, Univ. Nac. de La
Plata, Paseo del Bosque s/n, La Plata, Argentina\\
$^{2}$Instituto de Astrof\'isica La Plata, UNLP-CONICET, Paseo del Bosque s/n, La Plata, Argentina}
\begin{document}

\date{}

%\pagerange{\pageref{firstpage}--\pageref{lastpage}} \pubyear{2011}

\maketitle

\label{firstpage}

\begin{abstract}

We propose a scenario for the formation of DA white dwarfs with very thin
helium buffers.  For these stars we explore the possible occurrence of
diffusion-induced CNO-flashes, during their early cooling
stage.  In order to obtain very thin helium buffers, we simulate the formation
of low mass remnants through an AGB final/late thermal pulse (AFTP/LTP
scenario).  Then we calculate the consequent white dwarf cooling evolution by
means of a consistent treatment of element diffusion and nuclear burning.

Based on physically sounding white dwarf models, we find that the
range of helium buffer masses for these diffusion-induced novas to
occur is significantly smaller than that predicted by the only
previous study of this scenario. As a matter of fact, we find that
these flashes do occur only in some low-mass ($M\lesssim 0.6$\msun)
and low metallicity ($Z_{\rm ZAMS}\lesssim 0.001$) remnants about
$10^6 - 10^7$ yr after departing from the AGB.  For these objects, we
expect the luminosity to increase by about 4 orders of magnitude in
less than a decade.  We also show that diffusion-induced novas should
display a very typical eruption lightcurve, with an increase of about
 a few magnitudes per year before reaching a maximum of $M_V \sim
$ -5 to -6. Our simulations show that surface abundances after the
outburst are characterized by $\log N_{\rm H}/N_{\rm He}\sim
-0.15...0.6$ and N$>$C$\gtrsim$O by mass fractions. Contrary to
previous speculations we show that these events are not recurrent and
do not change substantially the final H-content of the cool (DA) white
dwarf.

Finally, with the aid of model predictions we discuss the possibility that
  Nova Vul 1670 (CK Vul) and the recently proposed [WN/WC]-central stars of
  planetary nebulae could be observational counterparts of this
  diffusion-induced nova scenario. We conclude that, despite discrepancies
  with observations, the scenario offers one of the best available
  explanations for CK Vul and, with minor modifications, explains the observed
  properties of [WN/WC]-central stars of planetary nebulae.

\end{abstract}


\section{Introduction}

It is well known that element diffusion plays a major role in shaping the
chemical stratification of white dwarfs \citep{1990RPPh...53..837K}.
Gravitational settling is responsible for the chemical purity of the outer
layers of white dwarfs. Due to the very high surface gravities of these stars
the lightest element remaining in the white dwarf (either H or He) floats up
and forms a pure envelope, the thickness of which gradually increases as the
white dwarf evolves. On the other hand, the large chemical gradients formed
during the previous evolution, lead to chemical diffusion which, in turn, will
smooth out the chemical profile of cool white dwarfs
(e.g. \citealt{2010ApJ...717..183R}).  Recently, the gravitational settling of
$^{22}$Ne in the liquid interior of white dwarfs, has been found to be a key
ingredient to explain the cooling age of the old, metal-rich, open cluster NGC
6791 (\citealt{2010ApJ...719..612A}, \citealt{2010Natur.465..194G}).

Since the first simulations of white dwarf evolution that included a
simultaneous treatment of diffusion and cooling
(i.e. \citealt{1985ApJ...296..540I, 1986ApJ...301..164I}) it was
noticed that diffusion could trigger thermonuclear CNO-flashes.  In
fact, the inward diffusion of H and the outward diffusion of C within
the pure He zone that is located below the H-rich envelope and above
the C-rich intershell (usually named ``He-buffer'', see
Fig. \ref{fig:scenario}) can lead to a runaway CNO-burning.  This
possibility was shown by \cite{1986ApJ...301..164I} who found that, if
the He-buffer was thin enough, C and H can come into contact and
develop an unstable H-burning.  This sudden energy release produces a
very rapid expansion, of the order of years, of the outer layers of
the white dwarf pushing the star back to a giant configuration and
increasing its visual magnitude from $M_V\sim 9$ to $M_V\sim -6$ in a
few years.  We term this eruptive event as ``diffusion-induced
nova''\footnote{\cite{1986ApJ...301..164I} named these events
  self-induced novas, but we will use the term ``diffusion-induced
  novas'' to distinguish them from late helium flashes (born again AGB
  stars, \citealt{1984ApJ...277..333I}) which predict similar
  lightcurves and can also be considered ``self-induced''.} (DIN)
although it leads to a much slower brightening than classical
novas. \cite{1986ApJ...301..164I} showed that after such events the
stars will become, in a few years, yellow giants with mildly
He-enriched surface compositions. In a more speculative mood they also
suggested that DINs may be recurrent, finally leading to H-deficient
compositions. Later, prompted by this speculation,
\cite{1990ARA&A..28..139D} suggested that the H-rich envelope could be
strongly reduced during these events, leading to DA white dwarfs with
thin H-envelopes, as inferred in some DA white dwarfs
(\citealt{2009MNRAS.396.1709C}).

Since the first exploratory simulation performed by
\cite{1986ApJ...301..164I}, these DINs have been mentioned in several review
papers and books (\citealt{1990ARA&A..28..139D},
\citealt{1995PhR...250....2I}, \citealt{2004PhR...399....1H},
\citealt{2005essp.book.....S} and more recently
\citealt{2010A&ARv.tmp....8A}), but no new numerical experiments have been
performed.  The latter becomes particularly relevant in light of the
criticism made by \cite{1986ApJ...308..706M} that the pre-white
dwarf models adopted by \cite{1985ApJ...296..540I} and
\cite{1986ApJ...301..164I} were not realistic enough. In this context, a
reexamination of the possibility and characteristics of DINs is in order.

As mentioned, DINs are expected to increase their visual magnitudes from
$M_V\sim 9$ to $M_V\lesssim -5$ in a few years. \cite{1986ApJ...301..164I}
mentioned that such event would display a lightcurve similar to that of a very
slow nova. The most famous object with such description is the enigmatic Nova
Vul 1670 (CK Vul), the oldest catalogued nova variable
(\citealt{2000PASP..112....1T}).  First discovered by P\`ere Dom Anthelme the
20th of June of 1670, and later independently by Johannes Hevelius on the 25th
of July, seem to have raised to 3rd magnitude in less than a year
(\citealt{1985ApJ...294..271S}), although no pre-maximum observations
exist. Its eruption lasted for two years (1670-1672). During that period it
faded and brightened two times before definitely disappearing from view. Three
centuries later, \cite{1982ApJ...258L..41S} partially recovered the
remnant. They discovered several nebulosities that were conclusively linked to
the 1670 eruption by \cite{2007MNRAS.378.1298H} through proper motion
studies. CK Vul remains a very enigmatic object and several scenarios have
been proposed to explain its observed behaviour. In fact, it has alternatively
proposed to be an hibernating nova, a very late thermal pulse (VLTP), a
``gentle'' supernova, a light nova and a stellar merger event
(\citealt{2007MNRAS.378.1298H}, \citealt{2003A&A...399..695K}). None of the
proposed scenario seems to correctly describe the observations
(\citealt{2007MNRAS.378.1298H}), although the VLTP seems to be the preferred
one (\citealt{1996PASP..108.1112H}, \citealt{2002MNRAS.332L..35E} and
\citealt{2007MNRAS.378.1298H}). In this context, and due to the similarities
between post-VLTP evolution and post-DIN evolution, it seems worth analysing
whether CK Vul could be understood within the DIN scenario.

 The main purpose of the present article is to study the possibility that DINs
 could take place in physically sounding white dwarf models with a realistic
 evolutionary history. We will also identify a detailed scenario for the
 creation of white dwarfs with thin enough He-buffers for DIN events to
 occur. Specifically, to perform this study we compute realistic white dwarf
 models by means of ``cradle to grave'' stellar evolution simulations. Then we
 compute white dwarf cooling sequences by considering a simultaneous treatment
 of element diffusion and evolution.

The article is organized as follows. In the next section we briefly describe
the input physics adopted in the present simulations. Then, in section 3, we
identify a possible evolutionary scenario that can lead to white dwarfs with
thin He-buffers. Once the evolutionary scenario has been set we present our
numerical simulations and results (Section 4). In particular, we show that the
proposed scenario can effectively produce white dwarf models with thin
He-buffers.  Then, we discuss the main predictions of the simulations and
explore under which conditions DINs might occur. Furthermore, we make a very
detailed description of the sequence of events before, during and after the
DIN events. Once the predictions of DIN simulations have been clarified, in
Section 5 we look for possible observational counterparts of these events. In
particular, we speculate that DINs could be the origin of CK Vul and PB8 (the
prototype of the recently proposed [WN/WC]-CSPNe, see
\citealt{2010A&A...515A..83T}) and show that, although significant
discrepancies exist, the DIN scenario offers a possible explanation for these
objects. Finally, we close the article with some final statements and
conclusions.

\section{Brief description of the stellar evolution code}

The simulations presented here have been performed with {\tt LPCODE} stellar
evolution code (see \citealt{2010ApJ...717..183R}). {\tt LPCODE} is a well
tested and amply used evolutionary code that is particularly appropriate to
compute realistic white dwarf structures and their evolution by means of
``cradle to grave'' stellar evolution simulations, and allows for a consistent
and detailed computation of the effects of element diffusion during the white
dwarf cooling phase.  As this work is aimed at exploring the outburst
properties of diffusion-induced CNO-flashes on hot white dwarfs, it is worth
mentioning that {\tt LPCODE} has been previously used to study other kind of
H-flashes on (very) hot white-dwarfs. These include the simulations of the
post-AGB born again scenario (\citealt{2007MNRAS.380..763M}), and the post-RGB
hot-flasher scenario (\citealt{2008A&A...491..253M}) for the formation of
H-deficient white dwarfs and subdwarfs respectively. Numerical and physical
aspects of {\tt LPCODE} can be found in \cite{2005A&A...435..631A}. In what
follows we only comment on the physical inputs regarding the treatment of
diffusion processes that are specifically relevant for the present work.

White dwarf models are obtained as the result of computing the evolution of
low mass stars from the ZAMS through the helium core flash and through the
thermal pulses on the AGB (TP-AGB) and, then, to the white dwarfs stage. {\tt
LPCODE} considers a simultaneous treatment of non-instantaneous mixing and
burning of elements, by means of a diffusion picture of convection coupled to
nuclear burning ---see \cite{2005A&A...435..631A} for numerical
procedures. The nuclear network considered in the present work accounts
explicitly for the following 16 elements: $^{1}$H, $^{2}$H, $^{3}$He,
$^{4}$He, $^{7}$Li, $^{7}$Be, $^{12}$C, $^{13}$C, $^{14}$N, $^{15}$N,
$^{16}$O, $^{17}$O, $^{18}$O, $^{19}$F, $^{20}$Ne and $^{22}$Ne, together with
34 thermonuclear reaction rates corresponding to the pp-chains, the CNO
bi-cycle, He-burning and C-ignition as described in
\cite{2006A&A...449..313M}.

As mentioned in the introduction, element diffusion strongly modifies
the chemical profiles throughout the outer layers of white dwarfs. In
this work white dwarf evolution was computed in a consistent way with
the changes of chemical abundances as a consequence of diffusion.  The
treatment is similar to that of \cite{1985ApJ...296..540I,
  1986ApJ...301..164I} but we consider, in addition  to
gravitational settling and chemical diffusion, the process of thermal
diffusion. We do not take into account radiative levitation, as it is
only relevant for determining surface chemical abundances and, thus,
is irrelevant for the purpose of the present work. Our treatment of
time-dependent element diffusion is based on the multicomponent gas
picture of \cite{1969fecg.book.....B}. Specifically, we solved the
diffusion equations within the numerical scheme described in
\cite{2003A&A...404..593A}.

\section{The evolutionary scenario}
\begin{figure}
\begin{center}
  \includegraphics[clip, width=8cm]{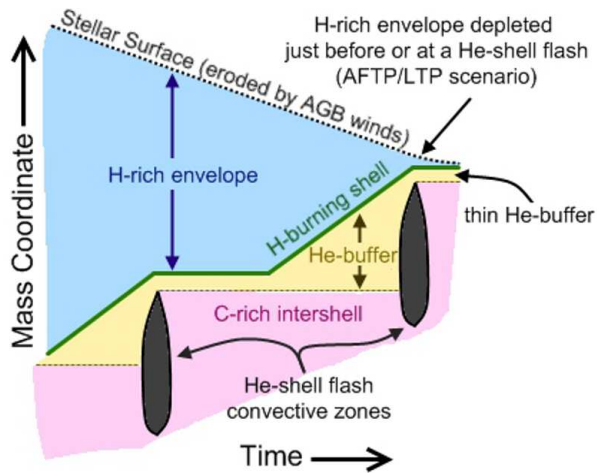}
\caption{Sketch of a Kippenhahn diagram of the proposed scenario for the
  formation of DA white dwarfs with thin He-buffers.}
\label{fig:scenario}
\end{center}
\end{figure}

\begin{figure*}
\begin{center}
  \includegraphics[clip, , width=14cm]{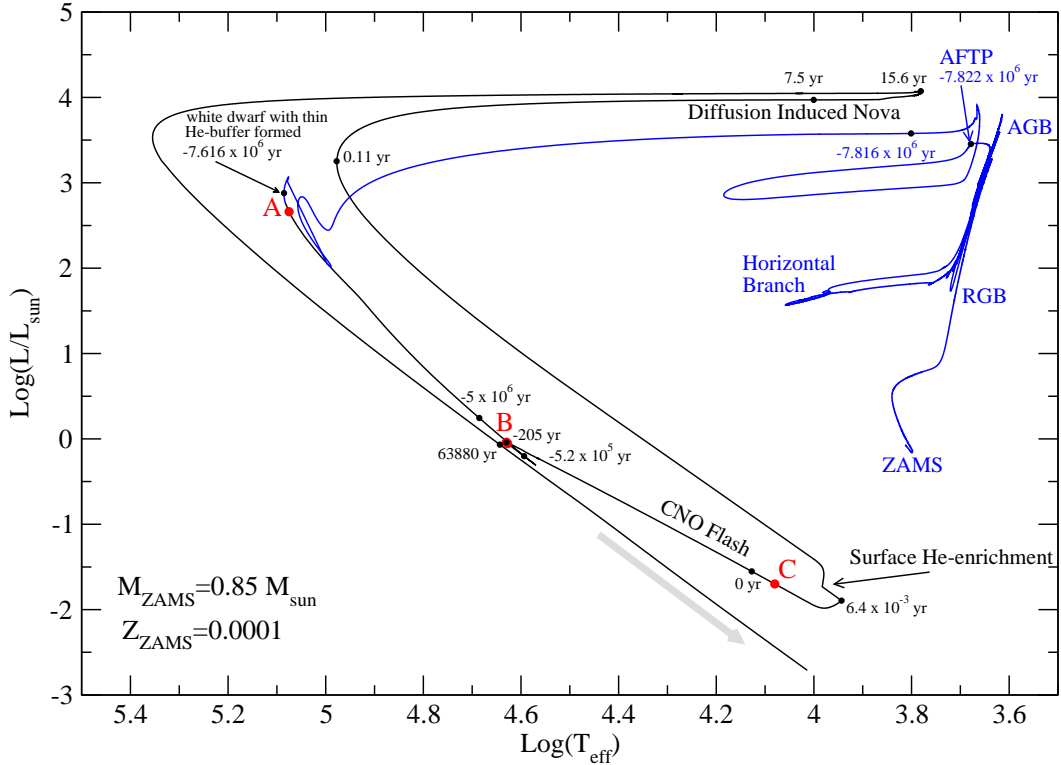}
\caption{HR-diagram evolution of an initially 0.85\msun\ star with
  $Z=1\times10^{-4}$. It undergoes its final thermal pulse on the AGB with a
  thin envelope and consequently no important H-reignition takes place before
  reaching the white dwarf cooling stage. This produces a white dwarf with a
  very thin He-buffer which experiences a DIN due to the
  diffusive mixing of C and H in the He-buffer (see
  Fig. \ref{fig:quimica}). Time has been choosen to be $t=0$ at the point of
  maximum energy release by H-burning during the CNO-flash.}
\label{fig:HR_ref}
\end{center}
\end{figure*}

The formation of a WD with a very thin He-buffer seems to be a key factor in
the occurence of a DIN episode (see \citealt{1986ApJ...301..164I}).
\cite{1986ApJ...301..164I} obtained a pre-white dwarf model with a thin
He-buffer by abstracting mass from an AGB star model shortly after it has
experienced a He-flash. \cite{1986ApJ...308..706M} noted that the way this
pre-white dwarf model was created is artificial and could affect the thermal
properties of the remnant. Whether white dwarfs with thin He-buffer can be
actually formed relies on identifying a scenario in which they could be formed
under standard assumptions.  As shown in Fig. \ref{fig:scenario}, after a
He-shell flash on the TP-AGB, the mass of the He-buffer region becomes
strongly reduced by intershell convection. Such thin He-buffer survives until
the reignition of the H-burning shell. Note that this may not be true if the
star experiences third dredge up events, during which the deepening of the
H-rich envelope will erase the remaining He-buffer by diluting it into the
envelope.  From this moment on, the He-buffer mass starts to increase until
the next He-shell flash. In this context, a realistic scenario that can lead
to a DA white dwarf with a thin He-buffer can be outlined. Recent studies on
the initial-final-mass relation (\citealt{2008ApJ...676..594K},
\citealt{2009ApJ...692.1013S}) support the absense of significant third dredge
up in low-mass stars (\citealt{2003...Karakas}).  In particular, the first
thermal pulses of low-mass stars ($M\lesssim 1.5M_\odot$) are not very strong
and, thus, no third dredge up takes place in numerical models. Hence, it is
not unreasonable to accept that low-mass stars experiencing an AGB final
thermal pulse (AFTP) or a Late Thermal Pulse (LTP) will end as DA white dwarfs
with thin He-buffers. In those cases, as no third dredge up happens, the very
thin He-buffer survives the last helium shell flash (either AFTP or
LTP). Then, during the He-burning phase that follows the flash, AGB winds will
erode an important fraction of the remaining (already depleted) H-rich
envelope, preventing a reignition of the H-burning shell and an increase in
the He buffer mass. As a result, the He-buffer is still very thin when the
star finally reaches the white dwarf phase. We propose that this scenario can
lead to white dwarfs with thin He-buffers and therefore might lead to DINs.

In order to bear out the scenario described in the preceeding
paragraph we have computed several sequences from the ZAMS to the
TP-AGB under standard assumptions ---see \cite{2010ApJ...717..183R}
for a description. Particularly, we have included overshooting only in
core convective zones within a diffusive convective picture as in
\cite{2006A&A...454..845M}. Thus, no overshooting mixing has been
included during the TP-AGB evolution of any sequence, as it leads to
strong third dredge up episodes in low-mass AGB stars.   Although
  the efficiency of third dredge up on AGB stars in clearly not a
  solved problem, it is clear that third dredge up is less efficient
  in low mass stars than in the more massive stars
  (e.g. \citealt{2002PASA...19..515K}). In this connection,
  \cite{2009ApJ...692.1013S} have shown that strong third dredge up is
  disfavoured, by the semiempirical initial final mass relation, for
  stars with masses lower than $4M_\odot$. It is worth noting that
  their study only deals with stars of masses $M>2 M_\odot$ and can
  not be directly applied to our work. However, as third dredge up
  efficiency is expected to decrease for stars of lower masses, it
  seems safe to assume that no third dredge up should be present in
  stars of even lower masses. In agreement with this,
  \cite{2006A&A...445.1069G} have suggested that the properties of
  carbon stars can be explained if no overshooting is added in low
  mass AGB stellar models. It must be noted, however, that both
  \cite{2009ApJ...692.1013S} and \cite{2006A&A...445.1069G} deal with
  stars of higher metallicities than studied in the present work and
  third dredge up is expected to increase at lower metallicities.
The absence of third dredge up episodes in low-mass AGB stars is of
special importance for the scenario outlined in the previous
paragraph, as third dredge up erodes the thin He-buffer formed just
after a thermal pulse.  It is worth noting that mass-loss rates during
the RGB and AGB phases have been adopted according to the
prescriptions of \cite{2005ApJ...630L..73S} and
\cite{1993ApJ...413..641V}.       This choice, together with the
  absence of third dredge up, is expected to produce a realistic
  initial final mass relation.   The only exception is mass loss just
before the final helium flash, which was tuned to obtain an AFTP/LTP
episode according to the scenario proposed here. It is worth
mentioning that standard AGB winds has been adopted during the
He-burning phase \emph{after} the last helium flash (AFTP/LTP), when
the star becomes a born again AGB star. No mass loss has been adopted
during the high temperature ($T_{\rm eff}$) phase of pre-white dwarf
evolution.

\section{Results}

In the previous section, we mentioned that several sequences have been
calculated for the present study. As all our DIN sequences display similar
characteristics, we will first describe the evolution of a choosen reference
sequence. Then, we will discuss how differences in progenitor/remnant masses,
initial metallicities and He-buffer thicknesses may affect the course of
evolution. In particular, we will show in later subsections the range in the
predicted abundances and timescales of all our computed sequences, as well as
the characteristics of those sequences that did not undergo DIN episodes.

\subsection{Detailed description of a reference sequence}
\begin{figure}
\begin{center}
  \includegraphics[clip, , width=8.5cm]{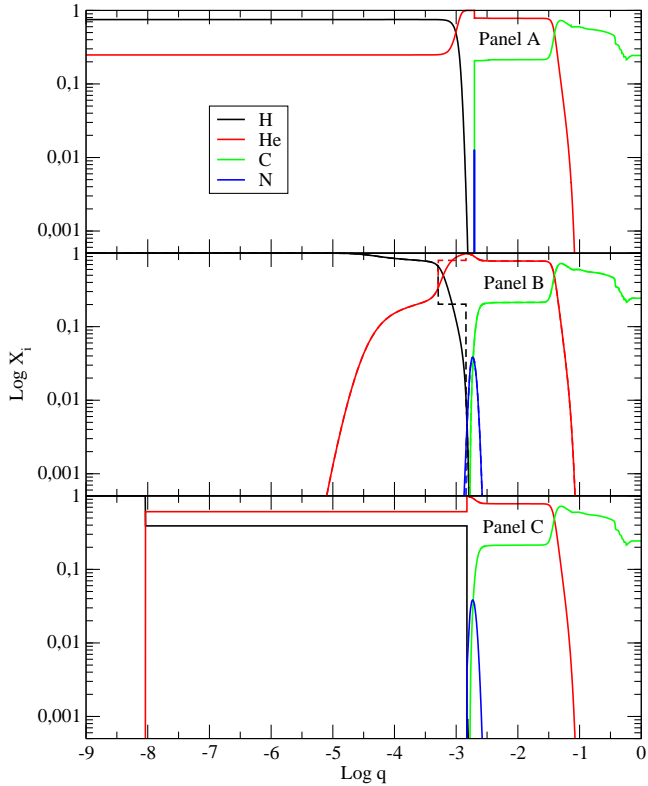}
\caption{Evolution of the chemical profiles of our reference sequence
  (0.53946\msun).  Panels A, B, and C correspond to the instants
  denoted by the same letters in Fig. \ref{fig:HR_ref}.  Dashed
    lines in the middle panel indicate profiles just after the
    development of the convective zone driven by the H-flash.}
\label{fig:quimica}
\end{center}
\end{figure}
\begin{figure}
\begin{center}
  \includegraphics[clip, , width=9.cm]{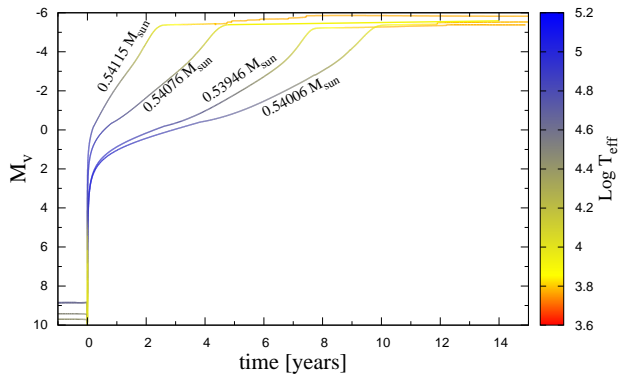}
\caption{Predicted lightcurves for the DINs of remnants
  coming from an initially $M=0.85$\msun, $Z=0.0001$ star. Different remnant
  masses were obtained by altering the wind intensity during the AGB final
  thermal pulse. Visual magnitudes were obtained from the predicted luminosity
  of the sequences and adopting the bolometric corrections derived by Flower
  (1996).}
\label{fig:curvas_1}
\end{center}
\end{figure}

\begin{figure}
\begin{center}
\includegraphics[clip, , width=8.5cm]{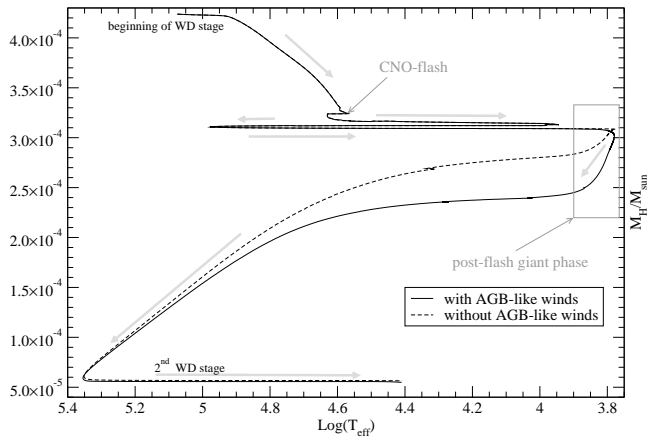}
\caption{Evolution of the total H-content of our reference white dwarf
sequence. The solid line shows the evolution in the case that AGB-like
winds are considered during the second AGB phase after the CNO-flash, while the
dashed line indicates the evolution of the H-content in the case no AGB-like
winds have been considered. Note that regardless the intensity of winds in
this stage the final H-content of the DA white dwarf is not modified and of
the order of $M_H\sim 5\times 10^{-5}$\msun}. 
\label{fig:Mdot}
\end{center}
\end{figure}

\begin{figure*}
\begin{center}
\includegraphics[clip, , width=17cm]{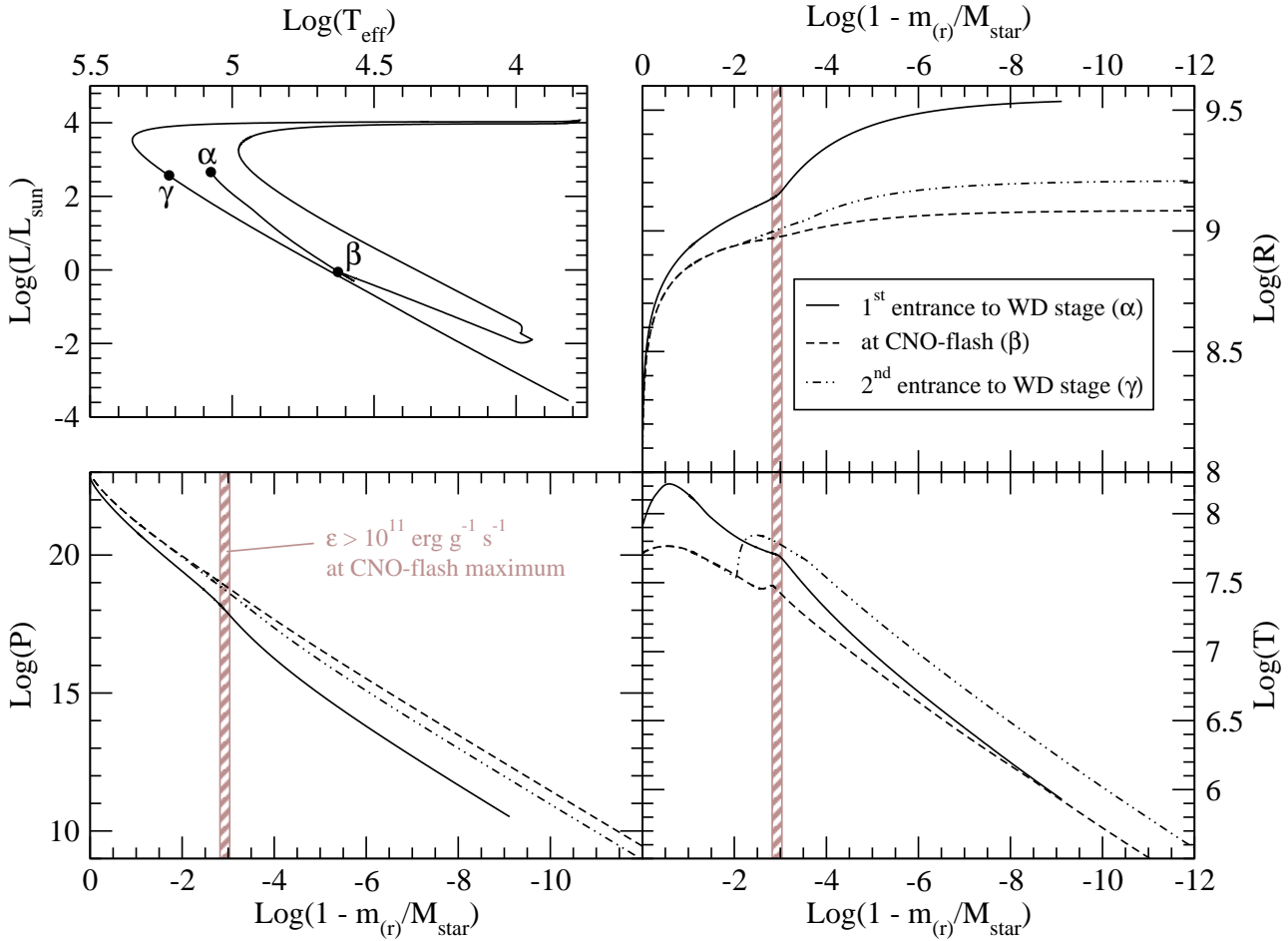}
\caption{Structure variables ($P(m)$, $T(m)$ and $r(m)$) at three different
  stages ($\alpha$, $\beta$ and $\gamma$ see upper left panel) of the white
  dwarf evolution. Note that the interior ($\log (1-m/M_\star)> -3$) structure
  of the white dwarf just before ($\beta$) and after ($\gamma$) the flash is
  almost unchanged. }
\label{fig:int_nova}
\end{center}
\end{figure*}

The evolution of our reference sequence in the HR-diagram is shown in
Fig.  \ref{fig:HR_ref}. It corresponds to an initially $M=0.85$
\msun\ (ZAMS) model star with $Z=0.0001$. After evolving through the
central H- and He- burning, the star starts climbing the early
AGB. This is so until the development of the first thermal pulse
($t_{\rm age}=11757.6$ Myr, $M=0.584$\msun), which marks the beginning
of the TP-AGB phase. The mass of the hydrogen-free core at that point
is $M_{\rm HFC}=0.515$\msun. Due to the low mass of the star at the
beginning of the TP-AGB, and owing to the relatively high winds
($\dot{M}\sim 10^{-7}$\msun/yr) the star suffers during the TP-AGB and
the long interpulse period ($P \sim 530000$ yr), the star experiences
only 3 thermal pulses. Mass loss during the very end of the TP-AGB was
tuned for the last thermal pulse to develop when the star was just
leaving the AGB (an AGB final thermal pulse; AFTP). The mass at the
moment of the AFTP is $M= 0.54403$\msun\ with a hydrogen-free core
mass of $M_{\rm HFC}= 0.53839$\msun. After winds have reduced the mass
of the star to 0.53946\msun, the star leaves the AGB ($T_{\rm
  eff}\gtrsim 10000$K) and evolves towards the white dwarf stage with
a H-envelope of $M_{\rm env}^{\rm PRE-WD}=1.07\times
10^{-3}$\msun,\footnote{We define the H-envelope as those regions with
  a H-content greater than $1\times 10^{-4}$ by mass fraction.}  which
corresponds to a total H-content of $M_{\rm H}^{\rm PRE-WD}=6.6\times
10^{-4}$\msun, and a He-buffer thickness of only $M_{\rm Buffer}^{\rm
  PRE-WD}=1.37\times 10^{-4}$\msun. Before entering the white dwarf
stage H is reignited, this produces the kink in the HR-diagram (see
Fig. \ref{fig:HR_ref}) just before the point of maximum $T_{\rm
  eff}$. As a consequence of this reignition, both the mass of the
H-envelope and the total H-content of the star are diminished to
$8.5\times 10^{-4}$\msun\ and $4.2\times 10^{-4}$\msun,
respectively. Conversely, the mass of the He-buffer zone is increased
up to $M_{\rm Buffer}=4.38\times 10^{-4}$\msun. At this point the
sequence begins its cooling as a white dwarf with a thin He-buffer see
(Fig. \ref{fig:HR_ref}). The chemical profile at that point is shown
in Fig. \ref{fig:quimica} (panel A).  From that point onwards, the
star cools and contracts as a white dwarf. Simultaneously, chemical
profiles are altered by element diffusion. In particular, H from the
envelope and C from the intershell region, just below the He-buffer
(see Fig. \ref{fig:scenario}), diffuse through the He-buffer due to
chemical diffusion triggered by the steep chemical profiles left by
the previous evolution.  Fig. \ref{fig:quimica} (panel B) shows the
chemical profile $7.58\times 10^6$ yr after the white dwarf started
its cooling evolution. At this point the amount of C and H diffused
into the He-buffer and, consequently the rate of energy generated by
CNO burning starts to increase. From this point onwards CNO burning
increases slowly until $\log L_H/L_\odot\sim 1.5$ when the rate of
energy generated by CNO burning leads to the development of convection
(about $23000$ yr after CNO burning started to increase). This
convective zone drags H from regions with higher H mass fractions and
thus leads to higher energy generation rates. As a consequence the
star enters a quicker phase and in less than 200 yr $L_H$ rises to
$\log L_H/L_\odot\sim 7.5$.  This sudden energy injection causes a
very quick expansion of the layers on top of the H-burning shell and
the star is balloned back to the giant branch in a few years (see
Fig. \ref{fig:HR_ref}). In Fig. \ref{fig:curvas_1} we show the
predicted visual lightcurve for the outburst episode ---$t=0$ denotes
the instant of maximum energy release by the CNO-flash, panel C of
Fig. \ref{fig:quimica}, see also Fig. \ref{fig:HR_ref}. Visual
magnitudes have been obtained from the predicted luminosity of the
sequences and adopting the bolometric corrections derived by Flower
(1996) for normal H-rich abundances. As can be seen after suddenly
rising from $M_V\sim 9$ to $M_V\sim 2$ the star increases its visual
magnitude much slower than a classical nova, at a speed of $\sim
1^{\rm m}/$yr. Specifically, the brightening speed just before
reaching its maximum (at $M_V\sim -5.2$, $\sim 8$ years after the
H-flash) is $ dM_V/dt\sim 1.6^{\rm m}/$yr.  Just after the maximum
energy release by the H-flash, when the star moves to low temperatures
and luminosities, the convective region driven by the H-flash merges
with the (very thin) external convective region caused by low
temperature opacity. Then, the material from the very outer layers of
the star is mixed with the material of the flash-driven convective
region. Consequently, the surface abundances of the model are
altered. The amount of He rises significantly ($X_{\rm He}= 0.61$ by
mass fraction) while the H mass fraction is reduced ($X_{\rm H}=
0.39$). Then, the DIN gives rise to an object with a mild
H-deficiency. Also N is increased at the surface of the star. As the
inner regions, which have experienced CNO burning, are N-enhanced, the
surface abundance of N during the outburst increases to $2.2\times
10^{-4}$, significantly higher than the mass fractions of C and O
($1\times10^{-5}$ and $1.1\times10^{-6}$, respectively), and higher
than the surface N abundance before the outburst ($N\sim 1.56\times
10^{-5}$).  Once back at low effective temperatures the star is
expected to undergo strong winds similar to those of AGB stars. The
time spent at low effective temperatures during this stage will be
strongly dependent on mass loss assumptions. If standard AGB mass loss
is assumed then the star reheates back to $T_{\rm eff}\gtrsim
10\,000$K in about $\sim 50$ yr, while in the case of no significant
mass loss the star stays at low effective temperature for about $\sim
260$ yr. Under the assumption of standard AGB mass loss, the total
mass lost by the object before it starts to contract again is of
$1.6\times 10^{-4}$\msun (see Fig. \ref{fig:Mdot}).
 Finally the star returns to the white dwarf cooling track. The model
returns to its pre-outburst stage ($\log L/L_\odot \sim -0.25$ and $\log
T_{\rm eff} \sim 4.6$) in only $\sim 10^5$ yr, much faster than in its
pre-outburst cooling, which took $\sim 7.6\times 10^6$ yr. This should not be a
surprise as only the outer layers are affected by the eruption and, so, the
interior of the white dwarf still resembles the structure prior to the
eruption (see Fig. \ref{fig:int_nova}). Only the outer layers, which have
much shorter thermal timescales, cool and contract during this second white
dwarf stage. The main consequence of this fast cooling is that diffusion has
not enough time to alter the composition of the inner regions of our new white
dwarf at high effective temperatures and the luminosity of the CNO-burning
shell remains very low. Then, no new CNO-flash does take place this second
time.

\subsection{Dependence on physical and numerical parameters}
In order to understand to what extent the outburst properties depend on the
remnant properties, we have performed several simulations under different
assumptions. Specifically, we have calculated the outburst for remnants of
different masses and also for remnants coming from progenitors of different
masses. We also explore how thin the He-buffer needs to be in order to obtain
a DIN. The latter is an important issue to understand how
frequent these events might be.

%Primero describir las secuencias de differentes masas provenientes de 
% igual progenitor que la de referencia (0.85)
To assess the importance of the He-buffer thickness for the development of
DIN, we have computed four additional sequences by shutting down winds in our
reference sequence at different times immediately after the maximum of
He-burning, during the final thermal pulse. Then, the remnant leaves the AGB
with thicker H-rich envelopes (and slightly different final masses). As a
consequence, when these remnants contract towards the white dwarf stage, the
reignition of H leads to thicker He-buffers. Hence,  we are able to follow the
cooling of white dwarfs with different He-buffer thicknesses (see Table
\ref{tab:0.85_WD}). Interestingly enough, the H-envelope mass of the
white dwarf is mostly dependent on the metallicity and not on the detailed
treatment of mass loss.
As shown in Table \ref{tab:0.85_WD} all the remnants derived from our
0.85\msun\ $Z=0.0001$ sequence, reach the white dwarf stage with very similar
H-rich envelopes (within a $\sim 10$\% difference in mass). This is because
the H-rich envelope of white dwarfs is mostly dependent on the mass of the
remnant and initial metallicity. On the other hand, remnants display
significantly different He-buffers thicknesses (with differences up to a
factor 5). As a consequence of this difference, the CNO-flash developes at
different points (luminosities) in the white dwarf cooling track. In fact, the
thicker the He-buffer, the longer time needed for diffusion to mix enough H
and C in the He-buffer and trigger the CNO-flash (see Table
\ref{tab:Nova_Prop}). Also, we find that in our 0.85 \msun\ $Z=0.0001$
sequence, if the He-buffer mass is $M_{\rm Buffer}\gtrsim 2.1\times
10^{-3}$\msun\ the sequence does not experience a DIN. In
Fig. \ref{fig:curvas_1} we show the lightcurves computed for these sequences.
In Table \ref{tab:Nova_Prop2} we show the expected abundances for major
chemical elements at the surface of the star during the nova outburst. As
mentioned in the previous section, the mixing of material from the He-buffer
with that at the H-rich envelope by convection, leads to an enhancement of the
surface abundances of He and N at the expenses of the H abundance. Depending
on the relative masses of the He-buffer and the H-rich envelope the outcome of
such mixing gives different values of H, He and N. In fact, for those
sequences with thicker He-buffers, the amount of He mixed during the
CNO-flash is higher. This can be clearly appreciated from both Tables
\ref{tab:0.85_WD} and \ref{tab:Nova_Prop}: the remnants that enter the white
dwarf stage with relatively thicker He-buffers are those who end with larger
He-abundances after the eruption. In all the cases, however, only a mild
($\log N_{\rm H}/N_{\rm He}\approx -0.15...0.6$) He-enrichment is attained.  It is worth
mentioning that the amount of H burnt during the nova event is not
significant and the star will ultimately end as a H-rich white dwarf (DA
white dwarf).

In order to explore to which extent the mass of the remnant and its
previous evolution would be relevant for the properties of DIN events,
we have calculated additional sequences for different progenitor
stellar masses. As in the case of our reference sequence, we have
evolved these sequences from the ZAMS up to the AGB phase and tuned
mass loss during the last thermal pulse in order to obtain a final
thermal pulse when the H-envelope mass was already very thin (i.e. an
AFTP/LTP). Again, as it happens with our reference sequence, (standard
AGB) winds during the AFTP/LTP erode a significant fraction of the
remaining H-envelope, inhibiting an important reignition of the
H-burning layer. Thus, white dwarfs with very thin He-buffers are
obtained. Namely, we have performed 3 additional cradle to grave
sequences with $Z=0.0001$ and ZAMS masses of $M=$ 1\msun;
1.25\msun\ and 1.8\msun. These sequences end as white dwarfs with
thin-He buffers and masses of 0.55156\msun\ ($M_{\rm H} =2.1 \times
10^{-4}$\msun; $M_{\rm Buffer} =3.6\times 10^{-3}$\msun),
0.59606\msun\ ($M_{\rm H} =2.8 \times 10^{-4}$\msun; $M_{\rm Buffer}
=9.8\times 10^{-5}$\msun) and 0.70492\msun ($M_{\rm H} =1.8\times
10^{-5}$\msun; $M_{\rm Buffer} =5.9\times 10^{-5}$\msun). While the
first two sequences suffered a DIN (see Table \ref{tab:Nova_Prop} for
their outburst properties), the last (more massive) one did not. In
fact in order to explore how robust this result was, we have followed
a similar approach than we did with our reference sequence. By 
  tuning the amount of mass lost during the very last AGB phase, we
obtained white dwarfs with very similar masses but significantly
different He-buffers and H-contents, namely: 0.70492\msun\ ($M_{\rm H}
=1.8\times 10^{-5}$\msun; $M_{\rm Buffer} =5.9\times 10^{-5}$\msun),
0.70496\msun\ ($M_{\rm H} =4.3\times 10^{-5}$\msun; $M_{\rm Buffer}
=5.9\times 10^{-5}$\msun), 0.70500\msun\ ($M_{\rm H} =7.5\times
10^{-5}$\msun; $M_{\rm Buffer} =8\times 10^{-5}$\msun),
0.70540\msun\ ($M_{\rm H} =6.2\times 10^{-5}$\msun; $M_{\rm Buffer}
=4.8\times 10^{-4}$\msun). In any of these sequences did we obtain a
DIN. Thus it seems that DINs do not occur in white dwarfs with masses
significantly higher than the canonical 0.6\msun\ value. To explore
further the importance of the white dwarf mass for the development of
DINs we computed additional sequences by artificially stripped mass
during the fifth thermal pulse of our
1.8\msun\ sequence\footnote{Under standard AGB winds the sequence
  experiences 17 thermal pulses before ending as a 0.70492\msun\ white
  dwarf.}  to obtain a 0.62361\msun\ white dwarf with a thin He-buffer
($M_{\rm H} =2.6\times 10^{-4}$\msun; $M_{\rm Buffer} =7.5\times
10^{-5}$\msun).  This sequence did experience a DIN and its properties
are described in Table \ref{tab:Nova_Prop}.

For the sake of completeness we have also performed two additional full
sequences with progenitors of $M=$ 1\msun\ but two different metallicities;
$Z=0.001$ and Z=0.01. These sequences end as white dwarfs of 0.55809\msun\
($M_{\rm H} =2.9\times 10^{-4}$\msun; $M_{\rm Buffer} =2.8\times 10^{-4}$\msun)
and 0.52382\msun\ ($M_{\rm H} =2.5\times 10^{-4}$\msun; $M_{\rm
Buffer} =5.5\times 10^{-4}$\msun). The first sequence did experience a nova
event (see Table \ref{tab:Nova_Prop}), while the second one did not. 

Although the exploration of the parameter space we have made is not completely
exhaustive, we can draw some preliminary conclusions. To begin with, it seems
that the occurrence of DINs is confined to low-mass, low-metallicity
remnants. This is true regardless the fact that our most massive remnant,
0.70492\msun, had a very thin He-buffer ($M_{\rm Buffer} =5.9\times
10^{-5}$\msun).  The absense of DINs in higher mass or metallicity remnants
with thin He-buffers is certainly not only because of the lower H-content that
higher mass/metallicity white dwarfs {\it are supposed} to have. 
In fact, the 0.52382\msun\ Z=0.01 sequence does not show significant
differences neither in mass nor in the values of $M_{\rm H}$ and $M_{\rm
Buffer}$ from that of our reference 0.53946 \msun\ Z=0.0001 sequence. The key
point that seems to prevent the DIN event in our more metallic sequence is
related to the luminosity of the CO-core. In fact, despite the slightly lower
mass and slightly higher carbon content, our 0.52382\msun\ Z=0.01 sequence
cools significantly faster ($\approx 50$\% at $\log T_{\rm eff}=4.6$) than our
reference 0.53946 \msun\ Z=0.0001 sequence. Consequently, the core luminosity
is roughly $\approx 50$\% higher than that of the reference sequence. This,
coupled to the fact that the CNO-burning shell is $\approx 25$\% less luminous
(at $\log T_{\rm eff}=4.6$) than the reference sequence case, makes
CNO-burning stable (see section \ref{sec:criterio}) and thus no DIN
developes. The faster evolution of higher metallicity warm white dwarfs was
shown in \cite{2010ApJ...717..183R}.  It seems, then, that higher metallicity
tends to prevent CNO-flashes, mostly due to a faster evolution during the warm
white dwarf stage which leads to a somewhat higher gravothermal energy release
in the core. In the case of our highest mass white dwarf ($\sim$
0.70500\msun), the absense of CNO flashes is related to the low luminosity of
the CNO-burning shell, which is, at most (at $\log T_{\rm eff}\sim 4.55$),
$L_{\rm CNO}\sim L_{\rm core}$ (see section \ref{sec:criterio}).  Then, our
numerical experiments, suggest that higher metallicity or masses tend to
prevent the occurence of DIN events due to the stabilizing effect of a low
$L_{\rm CNO}/L_{\rm core}$ value (see section \ref{sec:criterio}).

\begin{figure}
\begin{center}
  \includegraphics[clip, , width=9cm]{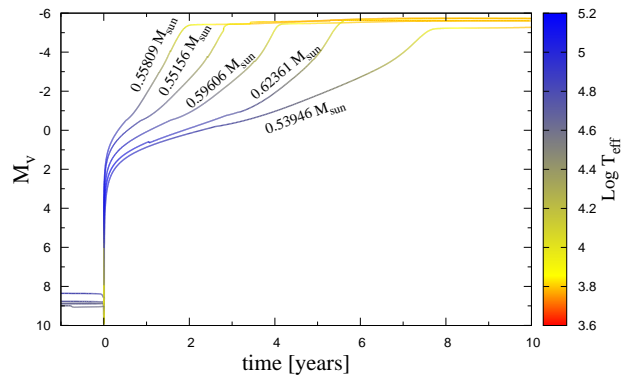}
\caption{Predicted lightcurves for the DINs of remnants with different masses
  and descending from different progenitors. See text for details.  Visual
  magnitudes were obtained from the predicted luminosity of the sequences and
  adopting the bolometric corrections derived by
  \citealt{1996ApJ...469..355F}.}
\label{fig:cur_gen}
\end{center}
\end{figure}

It must be noted that the final masses of our white dwarf remnants are in
agreement with our present understanding of the initial-final-mass
relationship for low/intermediate mass stars (\citealt{2009ApJ...692.1013S}),
and thus we expect our white dwarf models to be physically sounding.  In fig
\ref{fig:cur_gen}, we show the lightcurves and effective temperature evolution
during the outburst for sequences with different progenitors, metallicities
and final white dwarf masses.

Finally, our simulations show that DINs have characteristic properties that
make them easy to recognize.  First their visual lightcurve displays a sudden
increase, in less than a month, of 6 to 7 magnitudes followed by slower
increase in which it increases from $M_V\sim 2^m$ to its maximum of $M_V\sim
-5.5^m$ in a few years ---$\sim 2$ and $\sim 10$ yr depending on the remnant
properties. Its brightening speed just before reaching their maximum would be
of 1 to 4 magnitudes per year. Also, as the star brightness increases the star
should become cooler and yellowish (see Fig. \ref{fig:cur_gen}). Such
lightcurve would be probably classified as an extremely slow rising nova
(\citealt{1986ApJ...301..164I}). Secondly, the outburst abundances (see Table
\ref{tab:Nova_Prop2}) of all our sequences are characterized by $\log N_{\rm
H}/N_{\rm He}\sim -0.15...0.6$ and high N abundances of $10^{-4}-10^{-3}$ and
C$>$O, by mass fraction.

\subsection{The stability of CNO-burning on white dwarfs}
\label{sec:criterio}
\begin{figure}
\begin{center}
  \includegraphics[clip, , width=8.5cm]{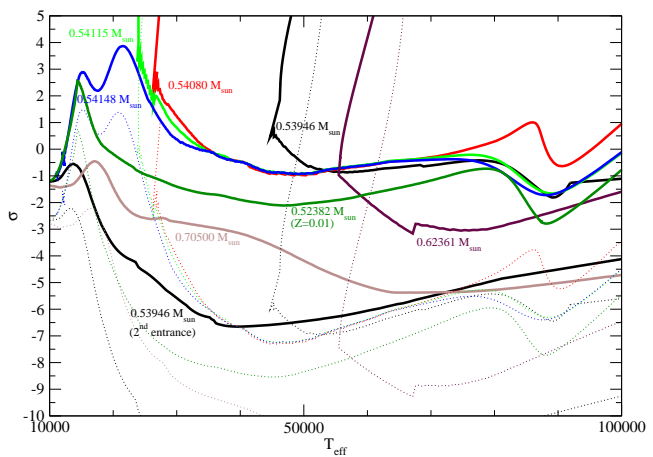}
\caption{Evolution of the value of the stability parameter $\sigma$ for
  selected white dwarf sequences. Note the significantly lower
  $\sigma$-values of the high mass (0.70500\msun) and high metallicity
  (0.52382\msun, Z=0.01) sequences which denote the stability of CNO-burning
  for those sequences. Also note the very low $\sigma$-value of our reference
  sequence during its second evolution of the white dwarf stage.}
\label{fig:sigma_teff}
\end{center}
\end{figure}
In this section, we will try to clarify the reasons beneath the instability of
the CNO-burning shell by looking for a  quantitative measure of the
stability of H-shell burning. In this connection, we have closely followed the
criterion presented by \cite{2004A&A...425..207Y} and extended it to be used
with mild burning shells in white dwarfs.

\cite{2004A&A...425..207Y} analysed the stability of a burning shell by
investigating the reaction of its mean properties under a uniform expansion of
the shell and the surrounding envelope. Specifically, they defined the burning
shell (where mean quantities are calculated) as those layers where the energy
generation rate is $\epsilon_{\rm CNO}>\epsilon_{\rm CNO\ at\ shell\
peak}/0.002$.  By assuming that the inner boundary of the burning shell
($r_0$) remains constant, neglecting the gravitational energy release of the
unperturbed model, and neglecting both the luminosity of the core ($l_0=0$)
and its perturbation ($\delta l_0=0$), \cite{2004A&A...425..207Y} found that
the temperature perturbation $\theta=\delta T/T$ follows the equation
\begin{equation}
\tau_{\rm th}\dot{\theta}=\sigma\theta.
\end{equation}
Where $\tau_{\rm th}$ is the thermal time scale of the shell and $\sigma$ is
given by
\begin{equation}
\sigma=\frac{\epsilon_T+ \kappa_T-4+
\frac{\alpha_T}{(\alpha_P\alpha_s-1)}
\left( \epsilon_\rho+ \alpha_s +\kappa_\rho \right)}
{1-\nabla_{\rm ad}\frac{\alpha_s\alpha_T}{(\alpha_P\alpha_s-1)}}.
\end{equation}
Where $\epsilon_T$, $\epsilon_\rho$, $\kappa_T$, $\kappa_\rho$ are the
logarithmic derivatives of $\epsilon (T,\rho)$ and $\kappa (T,\rho)$ while
$\alpha_T$, $\alpha_P$ are the logarithmic derivatives of $\rho
(T,P)$. $\alpha_s$ is a parameter describing the geometrical thickness of the
burning shell (see \citealt{2004A&A...425..207Y}).
Then, $\sigma$ is the parameter quantifying the shell instability, so that
$\sigma<0$ ($\sigma>0$) corresponds to stable (unstable) burning.

In our case, an important shortcoming of this approach comes from the fact
that the CNO-burning shell is not clearly detached from the outer pp-burning
shell. Thus, the choice of the borders of the simplified burning shell is not
as easy as in usual burning shells (see \citealt{2004A&A...425..207Y}). For
the present analysis we have choosen the lower boundary of the shell as the
point at which $\epsilon_{\rm CNO}=\epsilon_{\rm CNO\ at\ shell\ peak}/10$,
which is close to the point at which the structural effects of the burning
shell start to be apparent. For the outer boundary we have choosen it to be
the point closest to the shell maximum at which either $\epsilon_{\rm
CNO}=\epsilon_{\rm pp}$ or $\epsilon_{\rm CNO}=\epsilon_{\rm CNO\ at\ shell\
peak}/10$. The other important issue comes from the fact that the core
luminosity $l_0$ (see \citealt{2004A&A...425..207Y}) can not be neglected in
our case, due to the relative importance of the gravothermal energy release
from the core as compared to the shell luminosity ($l_{\rm CNO}=l_s-l_0$, where
$l_s$ stands for the luminosity at the outer boundary of the burning
shell). Also, the perturbation of the luminosity at the inner boundary of the
burning shell ($\delta l_0$) cannot be neglected either\footnote{Note that
this is also true for those cases in which $l_0\sim 0$.}. Within the scheme
presented by \cite{2004A&A...425..207Y} $\delta l_0$ can not be derived and,
thus, we will consider it to be proportional to $\delta l_s$. It is clear
that, for the reasons mentioned above, we can not expect from this approach to
get a quantitative prediction of the instability of the CNO-burning shell, but
we will show that its qualitative description is very useful to understand the
causes beneath the shell instability.

Under the assumptions and definitions mentioned above it is easy to show that
the criterion of \cite{2004A&A...425..207Y} is modified so that the parameter
$\sigma$ quantifying the shell instability is defined as
\begin{equation}
\sigma=\frac{\epsilon_T\, x+ \kappa_T\, y-4\, y+
\frac{\alpha_T}{(\alpha_P\alpha_s-1)}
\left( \epsilon_\rho\, x+ \alpha_s\, y +\kappa_\rho\, y\right)}
{1-\nabla_{\rm ad}\frac{\alpha_s\alpha_T}{(\alpha_P\alpha_s-1)}}.
\label{eq:sigma}
\end{equation}
 The main difference with the criterion presented by
\cite{2004A&A...425..207Y} comes from the presence of the parameters
$x=(l_s-l_0)/l_s$ and $y=(\delta l_s -\delta l_0)/\delta l_s$. For the present
analysis we have adopted extreme values for $y$; $y=1$ ($\delta l_0=0$, as in
\cite{2004A&A...425..207Y}) and $y=2$ (i.e. $\delta l_0= -\delta l_s$).

In all our sequences  $\alpha_s<0.16$ so the behaviour of the burning shell is not
far from that of $thin$ burning shells, where
\begin{equation}
\sigma_{\rm thin}\sim\epsilon_T\, x+ \kappa_T\, y-4\, y+
\alpha_T
\left( \epsilon_\rho\, x+\kappa_\rho\, y\right).
\label{eq:sigma_thin}
\end{equation}
In the absence of a core luminosity ($x=1$) $\sigma_{\rm thin}$, and thus
instability, is dominated by $\epsilon_T$. However, in our sequences we find
the factor $x$ to play a very important role as its values range from $x\sim
0.5$ for the 0.53946 \msun-sequence just before the flash to less than $0.2$
for the 0.70500\msun-sequence at its maximum value. An exploration of our
sequences shows that the term $\epsilon_T\, x$ is dominant in the behaviour of
$\sigma$.

In Fig. \ref{fig:sigma_teff} we show the evolution of the parameter $\sigma$
for some selected sequences. As expected, the quantitative prediction of the
instability is not reproduced. However qualitatively,
Fig. \ref{fig:sigma_teff} shows that instabilities develope during the rise in
the parameter $\sigma$ which is caused by the increase in $\epsilon_T$ due to
the cooling of the shell. However, this increase in $\epsilon_T$ is
counterbalanced by the decrease in $x$ as a consequence of the dimming of the
shell. The interplay between this two main factors is crucial for the
development of the instability. Diffusion leads to an increase of the
CNO-burning shell power, which increases $x$ and might trigger the flash. If,
by the time diffusion starts to change the H-content at the H-burning layers
the shell is already very dim H-burning remains stable and no DIN
develops. The existence of a thicker He-buffer delays the effect of diffusion
and thus, if the buffer is thick enough H-burning remains stable during the
white dwarf cooling. From Fig. \ref{fig:sigma_teff} we can also understand why
higher mass sequences ($M\sim 0.70500$\msun) do not undergo DIN events
despite the very thin helium buffers. The reason is that burning shells are
too dim as compared with the core luminosity and, thus,  very low $x$ values
render H-burning stable. In  the case of the high metallicity (Z=0.01)
0.52382\msun-sequence discussed in previous sections, its faster
cooling (than our reference Z=0.0001 sequences of similar mass) is tied to a
higher core luminosity which makes CNO-burning stable.

Thus, our analysis suggests that CNO-flashes depend on the intensity of the
CNO-burning shell (compared to the core luminosity, i.e. $x$) and its
temperature (i.e. $\epsilon_T$). Both factors depend on diffusion velocities,
which mixes H into deeper, hotter, layers increasing the CNO-burning power, and
the thickness of the He-buffer which delays the contact between C and
H. Finally, the relative intensity of the CNO-burning shell compared to the
core luminosity depends on the mass and (progenitor) metallicity of the white
dwarf, a fact that seems to prevent DIN events in higher mass and metallicity
remnants.

%decir los problemas, decir que se puede tratar como thin (alfa_s<0.16 lo que
%es muy bajito).

%decir que uno de los problemas es como definir la capa y el otro que la capa
%esta formada por 2 capas pegadas CNO y PP. 
%decir que a pesar de estos problemas puede usarse para entender el problema.
%decir que a diferencia de los shells de Langer delta l_O y l_o deben  tenerse
%en cuenta

\subsection{Effect of extramixing in the CNO-flash driven convective zone}
\label{sec:OV}
In the interests of completeness we have also analysed the impact of
extramixing in the flash-driven convective zone. In this connection, we have
recalculated the DIN event for our reference sequence including extramixing
beyond convective boundaries during the flash. This was done within the
picture developed by \cite{1997A&A...324L..81H} of an exponentially decaying
overshooting (OV) with $f=0.016$ (see \citealt{1997A&A...324L..81H} for a
definition). It is worth noting that, although the physical picture behind
this prescription might not be correct, it can be considered as an effective
way to include extramixing processes that take place at convective boundaries
(e.g. gravity waves; see \citealt{2007ASPC..378...43H}).

The main consequence of the inclusion of OV during the development of the
CNO-flash convective zone is that H is dragged deeper, into hotter regions of
the star. This enhances the CNO-burning luminosity during the flash, which can
be as high as $L_{\rm CNO}\sim 10^{10}L_{\odot}$, significantly higher than
the value of $L_{\rm CNO}\sim 10^{7}L_{\odot}$ predicted by no-OV
sequences. Also, due to the fact that the convective zone reaches deeper
during the flash, much more material from the N-rich He-buffer (and even from
the C-rich intershell) is dragged to the surface during the outburst. As a
consequence the surface abundances during the outburst are significantly
enhanced in He and N (see Table \ref{tab:Nova_Prop2}).  Also, surface C and O
are strongly increased to $\sim 10^{-3}$ (by mass fractions), as compared with
the $10^{-5}-10^{-6}$ resulting from the case when OV is neglected. It is
worth mentioning that the increase in C and O would be higher if the
intershell abundance of the model is richer in C and O (see section
\ref{sec:PB8} and Table \ref{tab:Nova_Prop2}). Also, as the CNO energy release
is orders of magnitude higher when OV is considered, the event is more
violent, reaching the giant stage in less than a year and showing a much
steeper outburst lightcurve (see Table \ref{tab:Nova_Prop}). Also, when OV is
included the duration of the giant stage ($\tau_3$, Table \ref{tab:Nova_Prop})
is also enhanced, being of the order of $\sim 10^4$ yr. This is because much
more H is consumed in this case and consequently much more energy is released,
allowing for a longer giant stage. Also as a consequence of this higher
H-burning, the amount of H remaining in the star by the time it reenters the
white dwarf stage is about $M_{\rm H}\sim 10^{-5}$\msun, six times smaller than in
the no-OV case ($M_{\rm H}\sim 5.7\times 10^{-5}$\msun)\footnote{Note that this
factor corresponds also to the difference between the duration of the giant
stage in the no-OV case, $\tau_3=260$ yr, and the OV case $\tau_3=1367$ yr.}

\subsection{Comparison with \citealt{1986ApJ...301..164I}}
%ACACACACACACA
In this section we compare our results with those of the DIN calculated by
\cite{1986ApJ...301..164I}. In Table \ref{tab:Nova_Prop} we show the main
characteristics of our simulations toghether with those extracted from
\cite{1986ApJ...301..164I}. An examination of \cite{1986ApJ...301..164I}
results show that all the predicted quantities fall within the range predicted
by our simulations. In particular, it is worth noting that the time,
luminosity and temperature at which the nova takes place in the simulation of
Iben \& MacDonald (1986; IM86) are very similar to those of our sequence of
similar metallicity (although different mass). The same is true for the
abundances of He and H during the outburst.  In view of the many numerical and
physical differences between \cite{1986ApJ...301..164I} and our simulations, we
consider the similarities of both results as a strong indication for the
robustness of the results and the feasibility of the scenario. In particular,
we feel confident that the results presented in Tables \ref{tab:Nova_Prop} and
\ref{tab:Nova_Prop2} reflect the main characteristics of DIN outburts.

Regarding the work by \cite{1986ApJ...301..164I} some comments need to be
done. The authors conclude that if the He-buffer thickness is smaller than a
critical value given by $M^{\rm CRIT}_{Buffer}\sim \Delta M_{H}/10 $, where
$\Delta M_{H}$ stands for the increase of the H-free core during the
interpulse phase on the AGB, then a DIN occurs. This conclusion has been
recently used by \cite{2006MNRAS.371..263L} to conclude that DINs can not
occur in white dwarfs descending from progenitors of masses lower than
3\msun. This result is completely at variance with the results presented in
the previous sections where all DINs come from low mass progenitors. However,
it should be mentioned that the condition $M^{\rm CRIT}_{Buffer}\sim \Delta
M_{H}/10 $ seems to be based on a single sequence, and thus it may be wrong to
extrapolate it to remnants of different masses. During the present study we
have not found such "critical value'' to be fulfilled.  In fact, in our 0.85
\msun\ sequences we find the frontier between the development or not of a
diffusion nova to take place somewhere between our 0.54148\msun\ ($M_{\rm
Buffer} =2.3\times 10^{-3}$\msun) and our 0.54148\msun\ ($M_{\rm
Buffer} =2.06\times 10^{-3}$\msun) remnant. For these sequences, the increase
of the H-free core between thermal pulses is $\Delta M_H=0.0101$\msun, very
similar to \cite{1986ApJ...301..164I} value. Then, the critical He-buffer
thickness is $M^{\rm CRIT}_{Buffer}\sim 2\times 10^{-3}\sim \Delta M_{H}/5
$. This is two times larger than suggested by \cite{1986ApJ...301..164I}. On
the other hand, for our 1.8\msun\ progenitor we find that while our
0.6236\msun\ ($M_{\rm Buffer} =7.5\times 10^{-5}$\msun) white dwarf remnant
does experience a CNO-flash, our slightly more massive remnant of 0.6239\msun\
($M_{\rm Buffer} =3\times 10^{-4}$\msun) does not. Then, as the increase of
the H-free Core for the 1.8\msun\ sequence during the fifth thermal pulse is
$\Delta M_H=0.009$, then the critical He-buffer thickness is $M^{\rm
CRIT}_{Buffer}\sim \Delta M_{H}/\alpha $ with $\alpha$ a number between 30 and
120.  Finally, let us mention that when we followed the AGB evolution of our
1.8\msun\ sequence up to 17$^{th}$ thermal pulse, the growth of the H-free
core during the interpulse phases is $\Delta M_H=0.0059$\msun. For the white
dwarf remnants created from this progenitor, even in the case with the thinner
He-buffer, 0.70492\msun\ ($M_{\rm Buffer} =5.9\times 10^{-5}$\msun), we did
not obtain a DIN. Even though in this case $M_{\rm Buffer} =\Delta M_H/100$,
no DIN developed. This is in line with our analysis of Section
\ref{sec:criterio} that suggests that for these white dwarf masses DIN events
will not develope, regardless of the He-buffer mass. Then, we conclude that
the previously mentioned critical He-buffer mass for the development of a DIN
is misleading. As mentioned in section \ref{sec:criterio} it is clear
that the He-buffer thickness is not the only ingredient that determines
whether a DIN will occur.

Finally, \cite{1986ApJ...301..164I} speculate\footnote{Due to numerical
difficulties they stopped their simulation once the star returned back to a
giant configuration.} that DINs might be recurrent. As mentioned in the
introduction, the occurence of recurrent DIN episodes has been invoqued
(\citealt{1990ARA&A..28..139D}) as a possible mechanism to explain the
existence of DA white dwarfs with thin H-envelopes (as those inferred by some
asteroseismological studies; e.g. \citealt{2009MNRAS.396.1709C}).  Our
computation of the aftermath of the DIN, shows that the star evolves much
faster during its second cooling stage as a white dwarf and thus, diffusion
has no time to significantly alter the chemical profiles of the inner regions
of the white dwarf. As a consequence, H-burning during the second white dwarf
evolution remains stable and no recurrent DIN events can develope.

\section{Possible observational counterparts}
\subsection{Speculations on the origin of CK Vul}
In the previous sections we have presented the predictions of the DIN
scenario. As these events have been suggested to look like ``very slow novas''
it seems natural to compare their predictions with some of those objects. The
most famous and enigmatic one is Hevelius nova CK Vul
(\citealt{1985ApJ...294..271S}). Here, we will compare the predictions of the
DIN with observations from CK Vul and with the predictions of the best
available scenario, the very late thermal pulse (VLTP;
\citealt{1996PASP..108.1112H}, \citealt{2002MNRAS.332L..35E} and
\citealt{2007MNRAS.378.1298H}).

\paragraph{Surface composition, H-abundance}
One of the most remarkable difference between VLTP models and observations of
CK Vul come from the significant H emission detected in the ejected material
of CK Vul, in contrast with the extreme H-deficiency detected in the ejected
material of the two bonafide VLTP objects; V4334 Sgr and V605 Aql
(\citealt{2007MNRAS.378.1298H}). In this connection, the DIN scenario predicts
a high H-content in the envelope during the ourburst and, thus of the ejected
material (see Table \ref{tab:Nova_Prop2}), contrary to the predictions of the
VLTP scenario (see \citealt{2007MNRAS.380..763M}). Also it is worth noting
that the most abundant CNO element predicted by the DIN scenario is N, which
is clearly detected in CK Vul surrounding material
(\citealt{2007MNRAS.378.1298H}).

\paragraph{Absence of an old planetary nebula} From the low dust emission
surrounding CK Vul, \cite{2007MNRAS.378.1298H} conclude that CK Vul must be an
evolved object $10^5-10^6$ yr older than the last time it experienced steady
high mass loss (e.g. on the TP-AGB). This time is orders of magnitude longer
than the time from the departure of the TP-AGB to a VLTP ($\sim 10^4$ yr). In
fact both V4334 Sgr and V605 Aql show old planetary nebulae surrounding the
new ejected material (\citealt{2007apn4.confE..17K}). On the contrary, the DIN
scenario predicts a much longer timescale between the cesation of high mass
loss and the time of the CNO-flash. As shown in Table \ref{tab:Nova_Prop},
this timescale is of the order of $10^6-10^7$ yr, in closer agreement with the
observations of CK Vul.

\paragraph{Ejected Mass} Due to the fact that the expanding layers in a DIN
episode are those above the region of the CNO-flash, DIN models have a very
stringent upper limit to the mass that can be expelled during the outburst. As
an example, the location of the H-flash in our reference sequence is $\lesssim
10^{-3}$\msun\ below the surface and the expanded envelope once the star is
back on the AGB is of about $\sim 10^{-4}$\msun. Thus, we can safetly conclude
that the ejected mass in the DIN scenario will be of $\sim 10^{-4}$\msun\
---and surely smaller than $\sim 10^{-3}$\msun. These values are within the
inferred masses for the ejected material in CK Vul, which is of
$10^{-5}$\msun\ - $5\times 10^{-2}$\msun\ depending on the assumptions
(\citealt{2007MNRAS.378.1298H}).

\paragraph{Outburst lightcurve}
 CK Vul was discovered at its maximum brightness on the 20th of June of 1670 as
3rd magnitude star. Although no description of the outburst lightcurve exist,
\cite{1985ApJ...294..271S} conclude that it is very unlikely that the star was
visible in the previous season. Thus, only a lower limit to the brightness
speed of $\sim 2$ magnitudes per year can be established. This limit is within
the values predicted by the DIN scenario (see Table
\ref{tab:Nova_Prop}). After reaching maximum brightness it faded and recovered
twice before finally disappearing from view. These fadings are remarkably
similar to those observed in born again AGB stars (V605 Aql;
\citealt{2002Ap&SS.279..183D} and V4334 Sgr; \citealt{2000AJ....119.2360D})
and RCrB stars (\citealt{1996PASP..108..225C}). In fact, this particular
lightcurve was one of the main arguments to link CK Vul to the VLTP scenario
(\citealt{1996PASP..108.1112H}, \citealt{2002MNRAS.332L..35E}). RCrB-like
fading are usually associated with carbon dust so one should wonder if such
fading episodes could take place in atmospheres which are not C-rich, as those
predicted by our DIN simulations (see Table \ref{tab:Nova_Prop}). In this
connection, we find suggestive that there is one ``Hot RCrB'', MV Sgr
(\citealt{2002AJ....123.3387D}), which has a carbon and nitrogen composition
very similar ($C \sim 5\times 10^{-4}$ and $N\sim 9\times 10^{-4} $) to those
predicted by the DIN scenario (N$>$C$>$O and $N \sim 10^{-3} - 10^{-4}$, see
Table \ref{tab:Nova_Prop2}), although MV Sgr shows a much stronger
He-enrichment. We consider that MV Sgr suggests that DIN events might also
display the fadings observed in RCrB like lightcurves, although we consider
this just an speculation and should be analysed in detail. From these
considerations we conclude that our present understanding of the DIN-scenario
is not in contradiction with CK Vul eruption lightcurve.

\paragraph{Maximum brightness} The maximum brightness is the main discrepancy
between observations and models (both for DIN or VLTP models). While
observations of CK Vul suggest at the preferred distance ${M_V}^{\rm CK
Vul}\sim -8$, DIN models predict significantly lower values of ${M_V}^{\rm
DIN}\sim -5.5$. It is worth noting, however that this is even worst for VLTP
models, in which fast-VLTP sequences predict ${M_V}^{\rm VLTP}\sim -4$
(according to the predicted luminosities in \citealt{2007MNRAS.380..763M}).

\paragraph{Actual luminosity of CK Vul} One of the
main discrepancies between inferences from CK Vul and both DIN and VLTP
scenarios comes from the radio emission. In fact, from the radio emission
detected by \cite{2007MNRAS.378.1298H} these authors derive a very low
luminosity of $\sim L_\odot$ for the ionizing object.  This is done by
considering that hydrogen is ionized by the photons emited by the central
object above the Lyman limit.  Neither the DIN nor VLTP events can account such
very low luminosities only 330 yr after the outburst. However, it is worth
noting that such very low luminosity is derived from the assumption that
the temperature of the central object is close to the effective temperature at
which the number of ionizing photons is maximum.  From the data presented by
\cite{2007MNRAS.378.1298H} and adopting the expression from
\cite{1994AJ....107.1338W},
\begin{equation}
F_c=1.761\times10^{48}\times a(\nu,T_e)\times \nu^{0.1}\times
{T_e}^{-0.45}\times S\times D^2,
\end{equation}
we derive that the number of ionizing photons is $N_\gamma \sim 4\times
10^{43}$ s$^{-1}$. As the number of ionizing photons is steeply dependent on
the effective temperature of the star, such number of photons can also be
attained by a star of $T_{\rm eff}\sim 10000 K$ and $L\sim 10000
L_\odot$. These values put the central object back in the AGB region of the
HR-diagram. Hence we are tempted to speculate that the radio emission could be
compatible with an object reheating back from the second stage in the AGB and
thus, compatible with both VLTP and DIN scenarios.  However, increasing the
luminosity of the star should increase the IR-flux from the obscuring material
(\citealt{2002MNRAS.332L..35E}). One may thus wonder if the IR-flux would be
much higher than the observational limits (e.g. $S_{12}\sim $mJy;
\citealt{2007MNRAS.378.1298H}). Some rough numbers show that this would be
strongly dependent on the geometrical and physical assumptions on the
obscuring material. For example, if we assume the star being obscured by a
circumstellar shell located at $r\sim 30$AU (and assumed to behave as a
blackbody), we obtain a shell temperature of $T_{\rm dust}\approx 720$K and a
flux of $S_{12}\sim$ 1000 Jy, several orders of magnitude above the
observational upper limit (\citealt{2007MNRAS.378.1298H}). On the contrary if
the star is obscured by the presence of a pre-existing edge-on disk ($r_{\rm
inner}=1$AU, $r_{\rm outer}=60$ AU and height $h=1$AU), as suggested by the
ejection almost perpendicular to the line of sight, then the IR flux would be
much lower. Indeed, by assuming a classical temperature profile ($T\propto
r^{-3/4}$; \citealt{1993ApJ...412..761N}) and a black body behaviour we
(roughly) estimate $S_{12}\sim 25$ mJy, much closer to the observational
constraints (\citealt{2007MNRAS.378.1298H}). 

We conclude that, if CK Vul is actually obscured by a thick disk, the
radio emission could be explained by a cool reborn giant star ($T_{\rm
  eff}\sim 10000 K$ and $L\sim 10000 L_\odot$) without an strong IR
emission. This would help to reconcile the predictions of both VLTP
and DIN scenarios with the observed properties of CK Vul. Then, a
detailed modeling of the dust surrouding CK Vul seems to be key in
order to reject a possible luminous cool central star as the present
state for CK Vul.

%It remains to be seen if such
%high luminosities could have some other measurable effects, although due to the
%short time past since the outburst, we wonder if most of the surrounding
%material could be adjusted to the new luminosity.

\subsection{Possible connection with PB8}
\label{sec:PB8}

As mentioned in Section \ref{sec:OV}, the inclusion of extramixing processes
at the boundaries of the CNO-flash driven convective zone leads to outburst
abundances rich in He, H and N (see Table \ref{tab:Nova_Prop2}). Also, the
inclusion of extramixing processes leads to a more violent flash and to an
increase in the time spent as a giant ($\tau_3$; see Table
\ref{tab:Nova_Prop}) after the DIN event. During this short giant stage the
reactivation of AGB winds is expected. Depending on the amount of mass lost
during this stage, it could be possible that the new ejected material might
form a new young planetary nebula once the star begins to contract and heat
towards the second white dwarf phase. Such young planetary nebula would show
very low dynamical ages and, together with its central star, a very atypical
He, H and N rich composition. PB8 has been recently shown to have
(qualitatively) similar characteristics (\citealt{2010A&A...515A..83T},
\citealt{2009A&A...496..139G}), namely low dynamical age of a few thousand
years and a He-, H- and N-rich composition. Specifically, the central star of
PB8 has been proposed as the prototype of [WN/WC], a new class of CSPN. In
particular, the surface abundances of PB8-CSPN have been derived, by
\cite{2010A&A...515A..83T}, to be [H/He/C/N/O]=[40/55/1.3/2/1.3].  These
abundances can be accounted by our DIN sequences that include extramixing
processes at convective boundaries ---see Table \ref{tab:Nova_Prop2}. Indeed,
as mentioned in Section \ref{sec:OV}, when overshooting is included at the
CNO-flash convective zone, convection drags material from deeper layers of the
white dwarf, reaching the upper layers of the C-rich intershell. Thus, the C
and O surface abundances after a DIN event will be strongly dependent on the
intershell composition of the star when departing from the AGB. In fact, if
intershell abundances are assumed to be similar to those of PG1159 stars (see
\citealt{2006PASP..118..183W}) much higher C and O abundances are obtained
after a recalculation of the DIN event (see Table \ref{tab:Nova_Prop2}). In
fact, the abundances obtained in this case are in good qualitative agreement
with those derived by \cite{2010A&A...515A..83T} for PB8-CSPN (i.e. He$\gtrsim
$H$>$N$\gtrsim$C$\sim$O). Also, the dynamical age derived by
\cite{2010A&A...515A..83T}, $\tau=2600$ yr, is qualitatively similar to the
giant phase duration of this sequence ($\tau_3\sim 1600$ yr). Taking into
account that our simulations correspond to a single, arbitrarily choosen
sequence, we find the similarities with the observed properties of PB8 to be
remarkable.  Further exploration of this possible explanation for PB8 seems
natural.

\section{Final remarks}

We have performed a detailed exploration of the diffusion-induced nova
scenario (DIN) originally proposed by \cite{1986ApJ...301..164I}. Our
calculations represent a significant improvement in our understanding of
CNO-flashes in warm white dwarfs.

Our main results can be summarized as follows:
\begin{itemize}
\item We have identified a definite scenario leading to the formation of
  DA white dwarfs with thin He-buffers. Such white dwarfs are naturally
  formed in low-mass stars, that do not experience third dredge up during the
  TP-AGB, and suffer from  either an AFTP or a LTP.
\item We have explored the parameter space of the DIN scenario and shown that
  there is a range of values of $M_{\star}$, $Z_{\rm ZAMS}$ and He-buffer
  masses for which DIN occur in physically sounding white dwarf models. Our
  results suggest that DINs take place in white dwarfs with $M_{\star}\lesssim
  0.6$ and $Z_{\rm ZAMS}\lesssim 0.001$ and thin He-buffers ---as those
  provided by the scenario described above.
\item Our simulations provide a very detailed description of the events
  before, during and after the DIN event. In particular, our results show that
  DIN events are not recurrent as previously speculated. Thus, DINs do not
  form  H-deficient white dwarfs, nor DA white dwarfs with thin
  H-envelopes. 
\item We have qualitatively described the mechanism by which the CNO-shell
  becomes unstable. Our analysis shows that the occurrence of CNO-flashes
  depends strongly on the intensity of the CNO-burning shell (as compared to
  the core luminosity), and its temperature. This seems to be in agreement
  with the fact that only our sequences with $M_{\star}\lesssim 0.6$ and
  $Z_{\rm ZAMS}\lesssim 0.001$ and thin He-buffers experienced DIN events.
\item Regarding the criterion presented by \cite{1986ApJ...301..164I} for the
  occurrence of DIN events, we find that such criterion is misleading as the
  He-buffer mass is not the only parameter that determines whether a DIN event
  will take place or not. In particular, for more massive remnants our
  simulations do not predict DIN for any possible He-buffer masses.
\item Our simulations provide a very detailed description of the expected
  surface abundances and lightcurves during the outburts. In particular we
  find that typical lightcurves display a maximum of $M_V\sim -5.5$, a
  brightness speed of a few magnitudes per year, and a mild He- and N-
  enrichment; with $N\sim 10^{-4}-10^{-3}$ by mass fraction and $\log N_{\rm
  H}/N_{\rm He}\sim -0.15...0.6$. Also, in all our sequences we find surfaces
  abundaces with $N>C>O$ by mass fraction.
\item We find that the inclusion of extramixing events at the boundaries of
  the CNO-flash driven convective zone leads to higher He, N, C and O
  abundances than in the case in which no extramixing is considered. Relative
  surface CNO abundances in these cases are $N>C\gtrsim O$ (by mass fractions),
  although the precise values will be strongly dependent on the C and O
  composition of the  He-, C- and O- rich intershell.
\item Finally, with the aid of our numerical simulations, we have discussed
  the possibility that Hevelius nova CK Vul and PB8-CSPN could be
  observational counterparts of DIN events. 

For CK Vul we find that despite discrepancies with observations the DIN
  scenario offers a possible explanation for CK Vul which is as good as the
  best available explanation to date (the VLTP scenario). In particular the
  DIN scenario can easily explain the absense of and old planetary nebulae
  surrounding CK Vul and the presence of H in the ejected material, which
  cannot be understood easily within the VLTP scenario.  Moreover, we suggest
  that the radio and IR emission from CK Vul (\citealt{2007MNRAS.378.1298H})
  could be explained if the star is assumed to be in its giant stage and
  surrounded by a pre-existing dust disk. Such reinterpretation of the
  observations would reconcile them with both the VLTP and DIN scenarios. A
  further exploration of this possibility seems desirable.

Regarding PB8, our simulations show that the atypical abundances observed in
  PB8 (\citealt{2010A&A...515A..83T}) and the low dynamical age of its nebula
  (\citealt{2009A&A...496..139G}) could be qualitatively understood within the
  DIN scenario if extramixing is allowed at the boundaries of the CNO-flash
  convective zone.
\end{itemize}

\section*{Acknowledgments}
M3B thanks P. Santamar\'ia and H. Viturro for
technical assistance. Part of this work has been supported through the grants 
PIP-112-200801-00904 from CONICET and PICT-2006-00504 from ANCyT.
This research has made an extensive use of NASA's Astrophysics Data System.
\bibliographystyle{mn2e}
\bibliography{mmiller}

\begin{landscape}

\begin{table}[P]
\begin{center}
\begin{tabular}{c|c c c c c c }
Remnant Mass  &  $M_{\rm env}^{\rm PRE-WD}$/\msun  &  $M_{\rm Buffer}^{\rm PRE-WD}$/\msun  &  $M_{\rm H}^{\rm PRE-WD}$/\msun  & $M_{\rm env} $/\msun  &  $M_{\rm Buffer} $/\msun  &  $M_{\rm H} $/\msun  \\\hline
0.53946\msun &   $1.07\times 10^{-3}$  &  $1.37\times 10^{-4}$  &  $6.7\times 10^{-4}$  &  $8.5\times 10^{-4}$  &  $4.38\times 10^{-4}$  & $4.2\times 10^{-4}$    \\
0.54006\msun &   $1.67\times 10^{-3}$  &  $1.37\times 10^{-4}$  &  $1.1\times 10^{-3}$  &  $8.9\times 10^{-4}$   &  $1.00\times 10^{-3}$   &  $4\times 10^{-4}$   \\
0.54076\msun &   $2.37\times 10^{-3}$  &  $1.4\times 10^{-4}$  &  $1.64\times 10^{-3}$  &  $9.9\times 10^{-4}$   &  $1.6\times 10^{-3}$   &  $4.9\times 10^{-4}$   \\
0.54115\msun &   $2.8\times 10^{-3}$  &  $1.4\times 10^{-4}$  &  $1.94\times 10^{-3}$  &  $9\times 10^{-4}$   &  $2.06\times 10^{-3}$   &  $4.7\times 10^{-4}$  \\
0.54148\msun$\dagger$ &   $3.0\times 10^{-3}$  &  $1.4\times 10^{-4}$  &   $2.15\times 10^{-3}$   &  $9\times 10^{-4}$   &  $2.30\times 10^{-3}$   &  $4.5\times 10^{-4}$   \\
\end{tabular}
\caption{Properties of post-AGB remnants of an initially
  0.85\msun\ star with $Z=1\times 10^{-4}$ which experiences a final
  thermal pulse (see text for details).  $M_{\rm env}$ is the mass of
  the outer layers with $X_H>10^{-4}$. $M_{\rm Buffer}$ Thickness of
  the almost pure He layers (He-buffer). $M_{\rm H}$ total hydrogen
  content of the remnant. $^{\rm PRE-WD}$ indicates values just after
  departing from the AGB, while the last three columns correspond to
  the same quantities at the begining of the white dwarf phase (after
  the H-reignition close to the knee in the HR-diagram). $\dagger$This
  sequence did not undergo a DIN, see text for details.}
\label{tab:0.85_WD}
\end{center}
\end{table}
%

%\begin{onecolumn}
\begin{table}[P]
\begin{center}
\begin{tabular}{c|c |c c c c c c c}
Progenitor Mass & White Dwarf Mass  & $\log L_{\rm pre}/L_\odot$  & $\log T_{\rm eff}^{\rm pre}$ &   $\tau_1$ [yr] & $\tau_2$ [yr]&  $|d{\rm M_V}/dt|$ &  $\tau_3$ [yr] & $\tau_4$ [yr]  \\\hline

%-----------------------------------------------------------------
0.85\msun & 0.53946\msun &  -0.31 & 4.57   & $7.6\times 10^6$   &  7.9   &1.6&260 &$\lesssim 1.6\times 10^5$\\
%0.85\msun & 0.53946\msun &  0.06 & 4.65   & $4.6\times 10^6$   &  7.9&1.6&260 &$\lesssim 8.9\times 10^4$\\
0.85\msun & 0.54006\msun$\dagger$ &  -0.46 & 4.54   & $1.2\times 10^7$   & 10.2   & 1.3 &-&-\\
0.85\msun & 0.54076\msun$\dagger$ &  -0.96 & 4.42   & $4\times 10^7$     &  4.9   &1.6  &-&-    \\
0.85\msun & 0.54115\msun$\dagger$ &  -1.19 & 4.37   & $7.7\times 10^7$   &2.7   & 2.4 &-&-\\
1\msun & 0.55156\msun &  -0.33 & 4.57   & $6.5\times 10^6$   &  3   & 3.3&270&$9.7\times 10^4$\\
1.25\msun & 0.59606\msun &  -0.05 & 4.65   & $3.4\times 10^6$   &  4.2   & 2.9 &216&      $5.5\times 10^4$\\
1.8\msun & 0.62361\msun$\ddagger$ &  0.25 & 4.73   & $2.2\times 10^6$    &  5.7   & 2.7 &49&$3.1\times 10^4$ \\
1.\msun ($Z=0.001$)& 0.55809\msun &  -0.47 & 4.54   & $8.9\times 10^6$  &  2.1& 3.9  &456&$1.1\times 10^5$ \\
0.85\msun & 0.53946\msun (w/OV)$\wr$&  -0.31 & 4.57   & $7.6\times 10^6$   &  0.66& 6.2 &1367 &$5.2\times 10^4$\\
0.85\msun & 0.53946\msun (w/OV, CO-rich)$\wr$,$\ast$&  -0.03 & 4.63   & $4.1\times 10^6$&  0.42& 3.1 &1628 &     $5.8\times 10^4$\\
IM86 ($Z=0.001$) & 0.6\msun &  -0.54 & 4.52   & $\sim 10^7$  &$\sim 7.5$   & -  & - & -   \\

%Agregar numero de pulsos termicos?
%
\end{tabular}
\caption{Outburst properties of the DINs studied in this work.  Unless
  stated otherwise, all sequences come from ZAMS progenitors with
  $Z=0.0001$. Timescales are defined as follows; cooling time at the
  moment of the CNO-flash ($\tau_1$), expansion time from the maximum
  energy release to the giant stage at $\log T_{\rm eff}=3.9$
  ($\tau_2$), duration of the cool ($\log T_{\rm eff}<4$) giant stage
  ($\tau_3$) and contraction time needed to reach the pre-outburst
  luminosity ($\tau_4$).  $\dagger$These remnants were obtained by
  reducing the mass lost during the final AGB thermal pulse from that
  predicted by standard AGB wind prescriptions. $\ddagger$This remnant
  was obtained by applying an artificially high wind during the fifth
  thermal pulse.$\wr$ These DINs were obtained form the
  0.53946\msun\ sequence by including overshooting (OV) in the
  convective zone generated during the CNO-flash. $\ast$ In this case
  the intershell composition of the remnant was modified to resemble
  the C- and O- rich surface abundances of PG1159 stars.  IM86
  indicates the sequence computed by Iben \& MacDonald
  \citep{1986ApJ...301..164I}.}
\label{tab:Nova_Prop}
\end{center}
\end{table}

\end{landscape}

\begin{onecolumn}
\begin{table}[P]
\begin{center}
\begin{tabular}{c|c | c c c c c }
Progenitor Mass & White Dwarf Mass  &  H & He & C &  N & O  \\\hline

%-----------------------------------------------------------------
0.85\msun    & 0.53946\msun             & 0.39  & 0.61 & $10^{-5}$         & $2.2\times 10^{-4}$ & $1.1\times 10^{-6}$      \\
0.85\msun    & 0.54006\msun            & 0.28  & 0.72 & $4\times 10^{-6}$ & $6.1\times 10^{-5}$   &   $3.6\times 10^{-7}$    \\
0.85\msun    & 0.54076\msun             &  0.19 & 0.81 & $10^{-5}$         & $1.4\times 10^{-4}$   & $10^{-6}$        \\
0.85\msun    & 0.54115\msun            &  0.15 & 0.85 & $4\times 10^{-5}$  & $3.3\times 10^{-4}$   & $3.2\times 10^{-6}$ \\
1\msun       & 0.55156\msun             &  0.38 & 0.62 & $2\times 10^{-4}$  & $4.2\times 10^{-3}$   & $ 10^{-6}$ \\
1.25\msun    & 0.59606\msun             &  0.49 & 0.51 & $5\times 10^{-5}$ &  $1.3\times 10^{-3}$   & $ 10^{-5}$  \\
1.8\msun     & 0.62361\msun             & 0.46 & 0.54 & $2\times 10^{-5}$ &  $4.6\times 10^{-3}$  &  $3\times 10^{-6}$    \\
1.\msun ($Z=0.001$)& 0.55809\msun       &  0.30 &  0.70 & $1\times 10^{-4}$ &$1.9\times 10^{-3}$  &  $4\times 10^{-6}$    \\
0.85\msun    & 0.53946\msun (w/OV)       & 0.22  & 0.73 & $6.6\times 10^{-3}$ & 0.038 & $1.1\times 10^{-3}$      \\
0.85\msun    & 0.53946\msun (w/OV, CO-rich) & 0.23  & 0.62 & 0.033  & 0.081& 0.033      \\
IM86 ($Z=0.001$) & 0.6\msun &   0.29 & 0.71 & $6.6\times 10^{-4}$ &  $7.5\times 10^{-4}$  &  $8.1\times 10^{-6}$    \\

%Agregar numero de pulsos termicos?
%
\end{tabular}
\caption{Outburst abundances of the DIN sequences presented in Table
  \ref{tab:Nova_Prop}.}
\label{tab:Nova_Prop2}
\end{center}
\end{table}

\end{onecolumn}
%\end{landscape}
%
\end{document}